\documentclass{aastex}
\usepackage{emulateapj5}
\usepackage{apjfonts}
\usepackage{natbib}
\usepackage{graphicx}
\usepackage{amsmath}

\newcommand{\lumin}{\mbox{$\rm\,ergs\,s^{-1}$}}
\newcommand{\gtsim}{\ensuremath{\gtrsim}}
\newcommand{\ltsim}{\ensuremath{\lesssim}}

\newcommand{\chisq}	{\mbox{$\rm\,\chi^2$}}
\newcommand{\amin}      {\mbox{$^{\prime}$}}
\newcommand{\aprx}	{\mbox{$\sim$}}
\newcommand{\rxte}      {{\em RXTE}}
\newcommand{\xray}	{\mbox{X-ray}}
\newcommand{\xrays}	{\mbox{X-rays}}
\newcommand{\hxt}{HEXTE}
\newcommand{\sax}{{\textit{Beppo}SAX}}
\newcommand{\eflux}	{\mbox{$\rm\,ergs~cm^{-2}~s^{-1}$}}

\def\ginga{\textit{Ginga}}

\def\degree{\hbox{$^\circ$}}
\def\cmsq{{$\rm cm^2$}}

\def\cgro{\textit{CGRO}}

\newcommand{\srcnm}	{XTE~J1946+274}

\shorttitle{Cyclotron Line in \srcnm}
\shortauthors{Heindl, et al.}

\begin{document}

\title{Discovery of a Cyclotron Resonance Scattering Feature in the X-ray Spectrum of \srcnm} 
\author{W.A. Heindl, W. Coburn, D.E. Gruber, R.E. Rothschild,}
\affil{Center for Astrophysics and Space Sciences, Code 0424, University of
California, San Diego, La Jolla, CA 92093}
\author{I. Kreykenbohm, J. Wilms, R. Staubert}
\affil{Institut f\"ur Astronomie und Astrophysik -- Astronomie,
University of T\"ubingen, Waldh\"auser Strasse 64, D-72076
T\"ubingen, Germany}

\email{wheindl@ucsd.edu}

\begin{abstract}
Observations of the transient accreting pulsar \srcnm\ made with the
\emph{Rossi X-ray Timing Explorer} during the course of the 1998
September--November outburst, reveal a cyclotron resonance scattering
feature (or ``cyclotron line'') in the hard \xray\ spectrum near
35\,keV.  We determine a centroid energy of $36.2_{-0.7}^{+0.5}$\,keV,
which implies a magnetic field strength of $3.1(1+z) \times
10^{12}$\,G, where $z$ is the gravitational redshift of the scattering
region.  The optical depth, $\tau = 0.33_{-0.06}^{+0.07}$, and width,
$\sigma = 3.37_{-0.75}^{+0.92}$\,keV, are typical of known cyclotron
lines in other pulsars.  This discovery makes \srcnm\ one of thirteen
pulsars with securely detected cyclotron lines resulting in
direct magnetic field measurements. 
\end{abstract}


\section{Introduction}

The transient accreting \xray\ pulsar \srcnm\ was discovered during an
outburst in 1998 September with the All Sky Monitor (ASM) on board the
\emph{Rossi X-ray Timing Explorer} (\rxte) \citep{Smi98}.  At the same
time, 15.8 second pulsations were observed with the \cgro/BATSE, with
the pulsating source designated GRO~J1944+26 \citep{Wil98}.  The best
\xray\ position \citep[\sax/MECS: 1\amin\ radius, 90\% confidence, ][]{Cam98}
falls within the {\textit{Ariel V}} error region for the
transient source 3A~1946+274, making it likely that the two objects
are identical \citep{Cam99}.

The initial outburst (see Fig.~\ref{f:asmlc}), which reached
\aprx6\,cps in the \rxte/ASM, lasted about 100\,days and was followed
by a series of smaller flares.  \citet{Cam99} discussed the extended
\rxte/ASM lightcurve, finding that the flaring was nearly periodic
with a repetition period of \aprx80\,d.  They interpret this as either
the half or full binary orbital period.  Either case is consistent
with the known orbital period range of Be/\xray\ binary pulsars.
Given its probable orbital period and the outburst characteristics,
\srcnm\ is most likely another example of a Be star/\xray\ binary
pulsar transient.  This class of binaries accounts for over half of
the known accreting pulsars \citep{Liu00}.

Several \rxte\ observations (see Figure~\ref{f:asmlc}) were made 
spanning the peak of the initial outburst.
In this \emph{Letter}, we report on the spectral analysis of these
observations and, in particular, the discovery of a cyclotron
resonance scattering feature, or ``cyclotron line'', at
\aprx35\,keV. \citet{San01} report their independent discovery of the
\srcnm\ cyclotron line with \sax\ in this volume.

Cyclotron lines result from the scattering of \xrays\ by electrons in
quantized Landau orbits in the \aprx10$^{12}$\,G fields near the
magnetic poles of accretion powered pulsars. The characteristic energy
of the Landau transition scales with the magnetic field as $E_{cyc} =
(11.6\text{\,keV})(1+z)^{-1}B_{12}$, where $E_{cyc}$ is the cyclotron
line energy in keV, $z$ is the gravitational redshift in the
scattering region, and $B_{12}$ is the magnetic field in units of
$10^{12}$\,G.  Because of this proportionality, cyclotron lines give
us the only direct measure of the neutron star magnetic field. In
general, harmonically spaced lines (corresponding to higher order
Landau transitions) may exist \citep[see for
example][]{Hei99,Hei00,San99}. Depending on the temperature and geometry of
the emitting and scattering material and the viewing angle with
respect to the magnetic field, the line profiles may be broad and
complex \citep[e.g.][]{Ara00,Kre00}.  To date, about a dozen pulsars
have well-established cyclotron lines, for the most part discovered
with \ginga, \rxte, and \sax\ \citep[see for
example][]{Mak99,Hei00,Dal00}.


\section{Observations} 

During the \srcnm\ outburst, 12 pointed observations were made with
the Proportional Counter Array (PCA) and the High Energy X-ray Timing
Experiment (HEXTE) on-board the \rxte\ (see Tab.~\ref{t:obs} and
Fig.~\ref{f:asmlc}). The PCA \citep{Jah96} is a set of five Xenon
proportional counter units (PCUs) sensitive in the energy range
2--60\,keV with a total effective area of \aprx 7000\,$\rm cm^2$. The
\hxt\ \citep{Rot98} consists of two arrays (``clusters A and B'') of 4
NaI(Tl)/CsI(Na) phoswich scintillation counters (15--250\,keV)
totaling \aprx1600\,\cmsq. Early in the mission, the pulse height
analyzer of a single cluster B phoswich failed, making the effective
area of cluster B approximately 3/4 of cluster A. The \hxt\ clusters
alternate pointing between target and nearby blank fields in order to
measure the background.  The PCA and \hxt\ fields of view are
collimated to the same 1\degree\ full width half maximum (FWHM)
region. In order to extend the life of the PCA detectors,
most observations are currently made with one or more PCUs turned off.
All the observations here were performed with PCUs 0, 1, and 2
on. Because PCUs 3 and 4 were sometimes off, they have been excluded
from this analysis.

\includegraphics[width=3.25in]{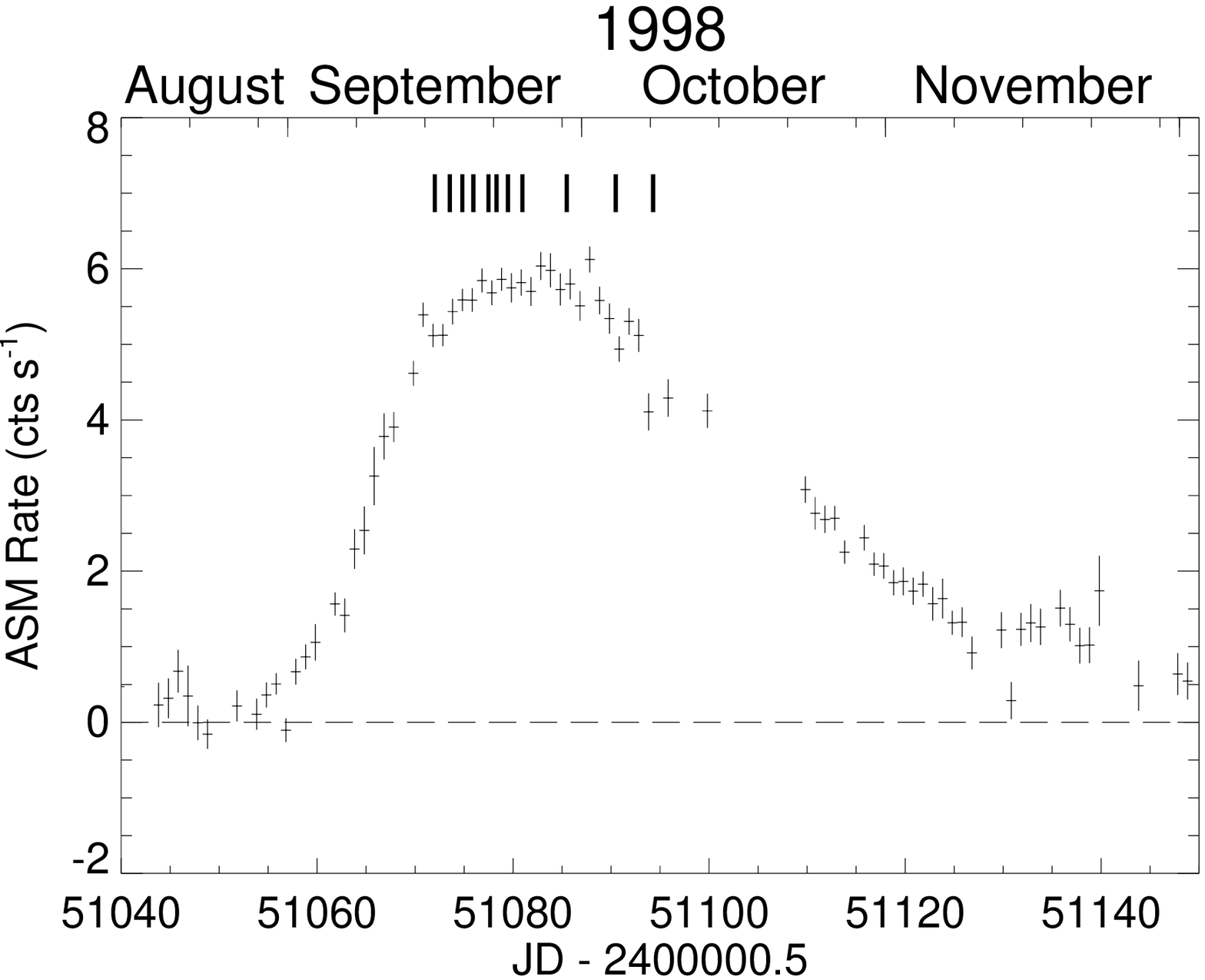}
\figcaption{The \rxte/ASM lightcurve for the 1998
September outburst of \srcnm.  Bars at the top indicate the PCA/HEXTE
observations (Table~\ref{t:obs}).}\label{f:asmlc} 

\section{Analysis}
For each observation, we accumulated source and background spectra
from the PCA standard 2 data and the \hxt\ event mode data.  We used
the FTOOLS 5.0.1 and the procedures detailed by \citet{Wil99}.
In order to take advantage of the best PCA calibrations, we limited
the data to detector layer 1 only.  We summed the spectra for the 3
active PCUs to form a single pulse height
spectrum.  We used the bright source background model (appropriate for
the counting rate of \srcnm) to estimate the PCA background.
Because of the high PCA counting rate of \srcnm\ and the duration
of the observations, the statistical errors in the energy spectra ($\ll
1\%$) were small compared to the uncertainties in the instrument
calibration.  It was therefore necessary to include systematic
uncertainties on the PCA data in order to achieve a reasonable fit to
the data.  We determined the size of these errors through the
procedures recommended by the PCA team \citep{Jah00} and explained in
\citet{Cob01}.  In short, we fit contemporaneous observations of the
Crab Nebula and Pulsar to a two power law model, using the \srcnm\
response matrices.  This model allows for different power law indices
for the pulsar and nebular components \citep{Kni82}.  We
adjusted the systematic errors until a reduced chi-squared of one was
achieved -- at a level of 0.3\% per channel below 20\,keV.  Above
20\,keV, the systematic errors became unacceptably large, \gtsim5\%,
so we chose to ignore these PCA data, rather than apply large systematic
errors.

\label{s:anal}
\begin{table*}[b]
\caption{\label{t:obs} Observations}
\begin{minipage}{\linewidth}
\renewcommand{\thefootnote}{\thempfootnote}
\begin{center}
\begin{tabular}{clccccc} \hline \hline
& & \multicolumn{2}{c}{PCA} & & \multicolumn{2}{c}{HEXTE}\\ 
\cline{3-4} \cline{6-7}
& Date\footnote{1998} & Time \footnote{exposure (ks)} & 
	Rate\footnote{cts s$^{-1}$ per PCU, 3--20\,keV}
	& & Time \footnote{exposure (ks)} & 
	Rate \footnote{cts s$^{-1}$, 16--100\,keV} \\ \hline
1  & Sep 16 & 1.42 & $195.0\pm0.2$ & & 0.84 & $30.1\pm0.5$ \\
2  & Sep 17 & 2.35 & $208.2\pm0.2$ & & 1.50 & $33.6\pm0.4$ \\
3  & Sep 18 & 2.51 & $217.5\pm0.2$ & & 1.64 & $34.6\pm0.3$ \\
4  & Sep 19 & 2.94 & $216.3\pm0.2$ & & 1.85 & $33.3\pm0.3$ \\
5  & Sep 21 & 2.69 & $250.0\pm0.2$ & & 1.76 & $37.9\pm0.4$ \\
6  & Sep 22 & 3.82 & $254.5\pm0.2$ & & 2.42 & $38.5\pm0.3$ \\
7  & Sep 23 & 2.70 & $243.0\pm0.2$ & & 1.75 & $37.6\pm0.4$ \\
8  & Sep 24 & 1.54 & $258.4\pm0.2$ & & 1.09 & $38.0\pm0.4$ \\
9  & Sep 29 & 4.93 & $244.9\pm0.4$ & & 3.26 & $36.4\pm0.2$ \\
10 & Oct 4   & 2.93 & $219.7\pm0.2$ & & 1.81 & $33.1\pm0.3$ \\
11 & Oct 8   & 2.77 & $200.6\pm0.2$ & & 1.79 & $30.7\pm0.4$ \\
12 & Oct 14  & 1.08 & $167.9\pm0.2$ & & 0.71 & $25.4\pm0.5$ \\
   & Total \footnote{Observations 1 -- 11. Observation 12 was excluded from the
                   analysis (see \S\ref{s:anal}).}            
		  & 30.6 & $230.7\pm0.3$ & & 19.7 & $35.3\pm0.1$\\ \hline \hline
\end{tabular}
\end{center}
\end{minipage}
\end{table*}

We accumulated the \hxt\ cluster A and B spectra separately then
summed them to form joint source and background spectra for each
observation.  For the analysis of these summed spectra, we added the
cluster A and B response matrices, weighted by the cluster effective
areas (4:3, respectively). Fits with this matrix to a single power law
model, appropriate in the \hxt\ energy band, to the Crab have
residuals of $\pm$1\%.  This is smaller than the statistical
uncertainties, so we applied no additional systematic errors to the
\hxt\ data.

Because the observations were spread over \aprx1\,month and a range of
\aprx40\% in flux, we were concerned that changes in the source
spectrum could affect our analysis.  To search for spectral changes,
we calculated for each observation the ratio of the net PCA and HEXTE counts
spectra to the total spectrum of all observations (see
Figure~\ref{f:ratio}). In all but observation 12, we 
\includegraphics[width=3.25in]{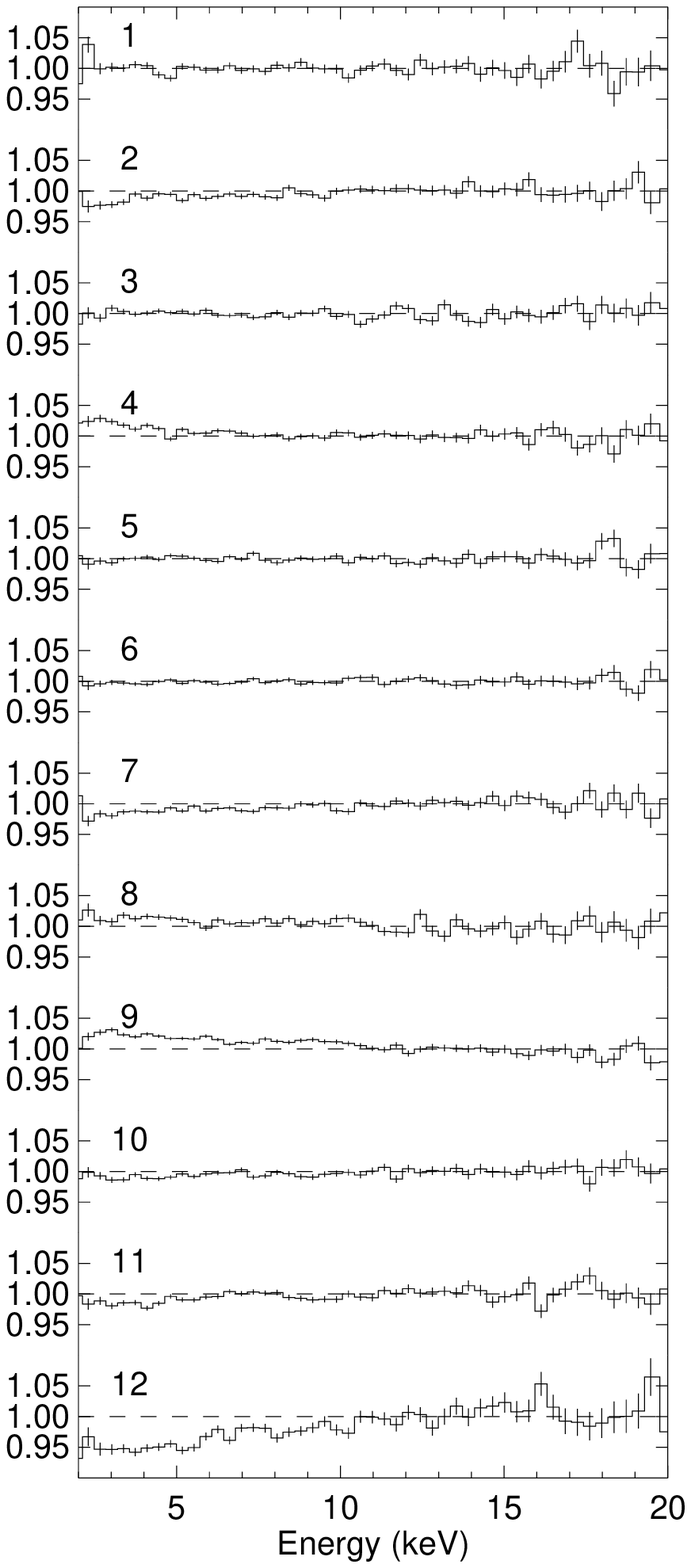}
\figcaption{Ratios of the net PCA counts spectra from the
twelve observations (Table~\ref{t:obs}) to the total spectrum,
normalized to the 8--20\,keV counting rate.  While a small residual
slope is apparent in observation 9, deviations above 8\,keV were less
than 2\%, so this spectrum was included in the analysis.  However,
owing to the overall steep slope of its ratio and large deviations from one,
observation 12 was eliminated.}\label{f:ratio} 

\vspace{1ex}
\noindent found that above8\,keV, the ratio was nearly flat, 
indicating that only the flux, and not the
spectral shape, had changed.  In observation 12 (the last and dimmest
observation -- see Table~\ref{t:obs}), significant spectral changes
were observed, and this pointing was eliminated from our analysis.
Observations 1 to 11 showed no changes in spectral shape at a level of
\aprx1--2\% between 8--20\,keV.  Below 8\,keV, systematic changes of
up to \aprx5\% were
\includegraphics[width=3.25in]{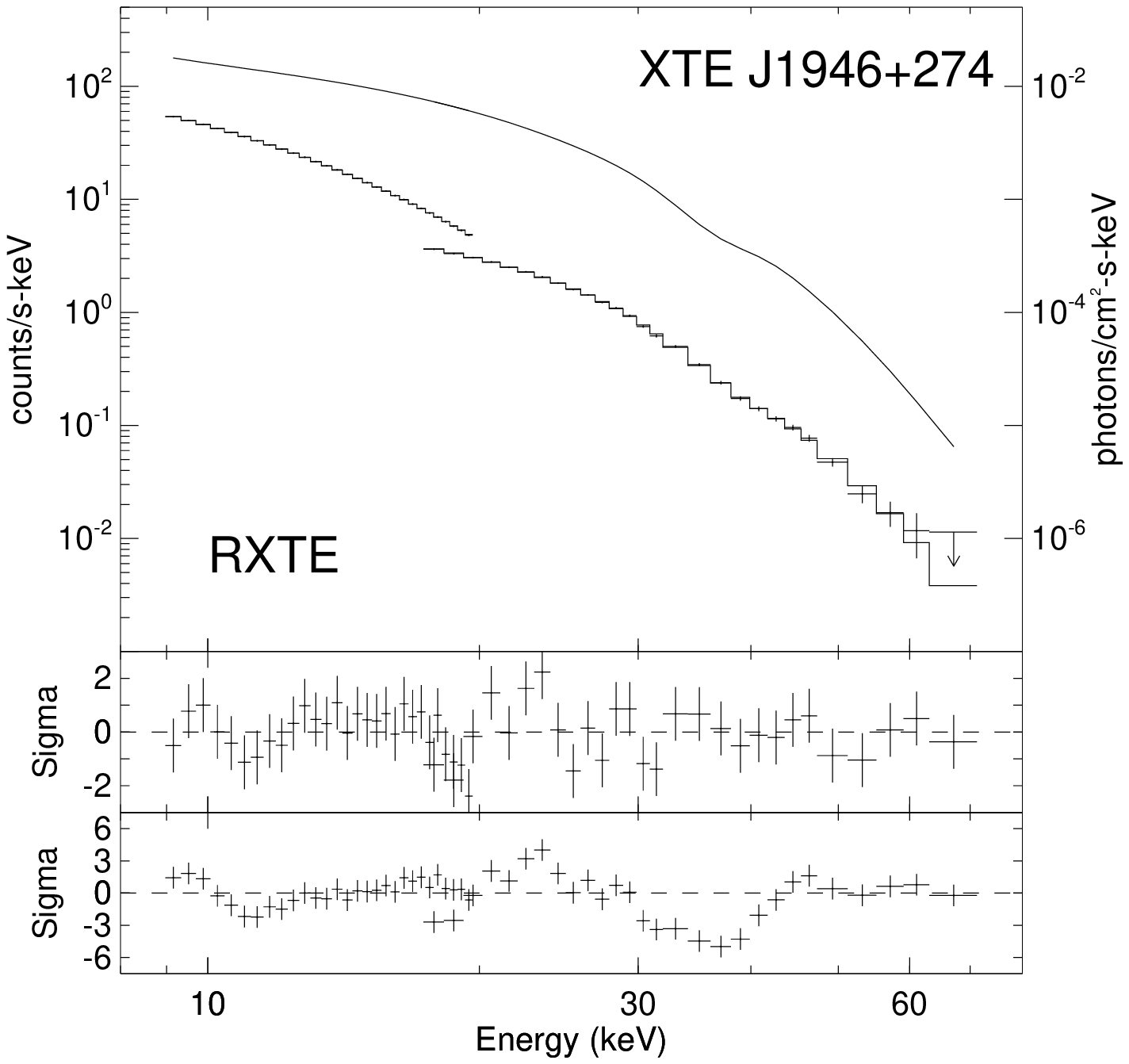}
\figcaption{The joint PCA/HEXTE spectrum of \srcnm.  Top
panel shows the data (crosses), the best fit NPEX model with a
cyclotron line (histograms), and the inferred incident photon spectrum
(smooth curve). See Table~\ref{t:fit} for fit parameters. The middle
and bottom panels show the residuals for the best fit model and the
best fit NPEX model with no cyclotron line.}\label{f:spec} 

\vspace{1ex}
\noindent seen, particularly in the lowest energy channels.
These changes are consistent with \ltsim10\% variations in the
absorbing column and the iron line emission. Such variations are not
unexpected during a Be/\xray\ binary outburst, and they are likely not
indicative of changes in the underlying continuum emission.  The HEXTE
ratios showed no evidence for spectral variability at a limiting level of
\aprx5--10\% below 35\,keV. Above 35\,keV the statistics did not allow
for a meaningful comparison. For these
reasons, we concluded that by excluding data below 8\,keV we could
safely combine observations 1 to 11.

After accumulating observations 1 through 11 as described above, we
attempted to fit the PCA and HEXTE spectra jointly with a set of
continuum models typically used for accreting pulsars.  While there is
a notable lack of adequate theoretical models, pulsar spectra are
heuristically well-described by a power-law at low energies which
breaks to an exponentially cut-off power-law at high energies
\citep{Whi83}.
\citet{Kre99} review several models with this asymptotic behavior
-- the exponentially cut-off power law \citep[PLCUT,][]{Whi83}, the
Fermi-Dirac cut-off \citep[FDCUT,][]{Tan86}, and a combination of two
power-laws with a positive and a negative exponent
\citep[NPEX,][]{Mih95}.  None of these models provided an acceptable
fit to the \srcnm\ spectrum. All models left significant, cyclotron
line-like residuals
near 35\,keV (see Figure~\ref{f:spec}).  We therefore added to each model a
cyclotron absorption term with a Gaussian optical depth profile: 
\begin{equation}
\tau(E) = \tau_{\text{cyc}} \times e^{-(E - E_{\text{cyc}})^2/{2\sigma_{\text{cyc}}^2}}.  
\end{equation}  
Adding the Gaussian absorption line greatly improved the fits.  The
best fit was achieved with the NPEX continuum, given by:
\begin{equation}
\text{NPEX}(E) \propto (E^{-\Gamma_1} + \alpha E^{+\Gamma_2}) \times
e^{-(E/E_{\text{fold}})} 
\end{equation}  
where $\text{NPEX}(E)$ is the photon flux at energy $E$; $\Gamma_1$
and $\Gamma_2$ are the indices of the falling and rising power law
components, with $\alpha$ their relative normalizations; and
$E_{\text{fold}}$ is the exponential folding energy.  Our best-fit
model was of the form $ \text{Flux} \propto \text{NPEX}(E) \cdot
\text{e}^{-\tau (E)}$.  In this case, the reduced $\chisq$ changed
(with the addition of the line) from 3.61 to 0.97 for 52 and 49
degrees of freedom respectively. The resulting $F$-Test probability
for this to be a chance improvement is $ 1.2 \times 10^{-14}
$. Table~\ref{t:fit} gives the best fit parameters for this model.
\newpage

\section{Results and Discussion}

Figure~\ref{f:spec} shows the best fit model and residuals together
with the inferred incident photon spectrum.  It also shows residuals
to the best NPEX model without a cyclotron line, which is then
apparent in the residuals around 35\,keV.  From the cyclotron line
energy, we deduce a magnetic field strength in the scattering region
of $3.1(1+z) \times 10^{12}$\,G. The line is weakly resolved with the
HEXTE energy resolution of \aprx8\,keV (FWHM, at 35\,keV), and the
fitted width of the line is $\rm 9.3^{+2.5}_{-2.1}$\% of the line
energy.  This is within the range of typical values determined with
the Gaussian absorption profile model \citep{Cob01b}.  No harmonic
line near 70\,keV was found, but as is evident in Figure~\ref{f:spec},
the falling high energy continuum provides inadequate statistics to
make a sensitive search for such a feature.
\vspace{2ex}

\includegraphics[width=3.25in]{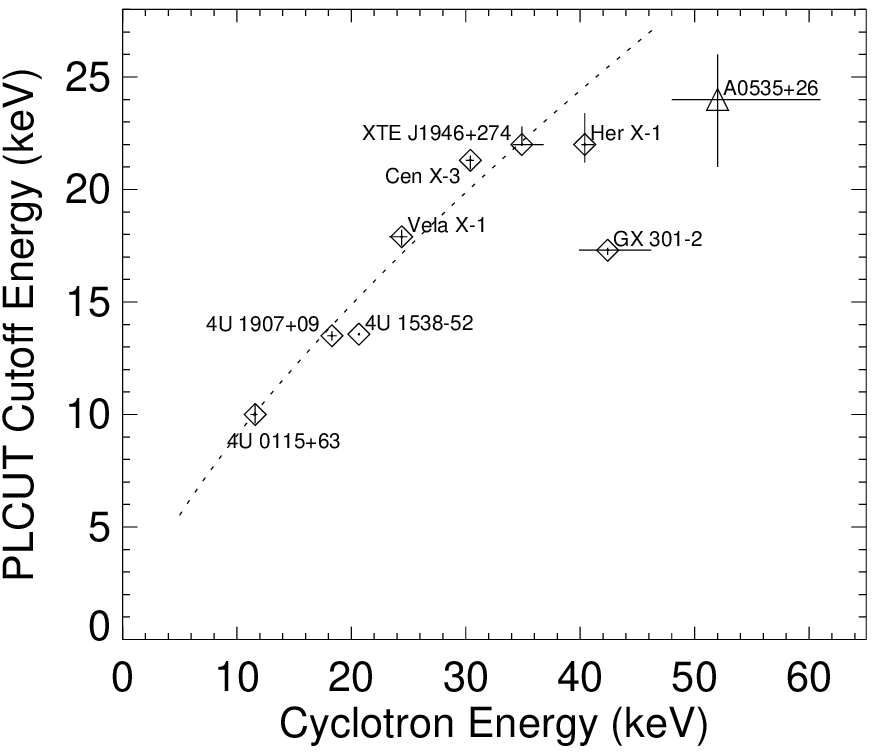}
\figcaption{Spectral cut-off energy as a function of
cyclotron line energy.  A possible saturation of the correlation is
seen at energies above 30\,keV.  All points are from \rxte\ except
A0535+26 which is from HEXE/TTM \citep{Ken94}.  The line is a power law
with $E_{cut} \propto E_{cyc}^{0.7}$ \citep{Mak99}.  See text for
parameter definitions.}\label{f:ecutvsecyc} 
\vspace{1ex}

Figure~\ref{f:ecutvsecyc} \citep[after][]{Cob01b} shows the spectral
cut-off energy plotted against the cyclotron line energy for 8 pulsars
measured with \rxte.  Also shown is the (somewhat controversial) HEXE/TTM
result for A0535+26 \citep{Ken94}. For uniformity, the pulse-phase
average spectra of all sources were fit with the PLCUT continuum:
\vspace{0.5ex}
\begin{equation}
\text{PLCUT}(E) \propto E^{-\Gamma} \times \left\{
	\begin{array}{ll}
		1 & \text{if~} E \leq E_{\text{cut}}\text{;} \\
		e^{-(E-E_{\text{cut}})/E_{\text{fold}}} &  \text{otherwise}
	\end{array}	
	\right.
\end{equation}  
smoothed at $E_{cut}$ \citep[see][~for details]{Cob01b}.  Two sources
are excluded from the plot: 4U~1626-67, whose cut-off energy varies by
a factor of four with pulse phase \citep{Cob01b}, and the low
luminosity (\aprx$4\times 10^{34}$\,\lumin) object 4U~0352+309 (X~Per)
whose spectrum is not well fit by the usual pulsar models. Due to the
complex shape of the 4U~0115+63 fundamental line \citep{Hei00}, its
plotted energy is half of the first harmonic value.

\begin{table*}[b]
\caption{Best fit spectral parameters for an
NPEX times Gaussian absorption profile model.}\label{t:fit} 
\begin{minipage}{\linewidth}
\renewcommand{\thefootnote}{\thempfootnote}
\begin{tabular}{ll} \hline \hline
Parameter & value \\ \hline
Continuum & \\ \cline{1-1}
$\Gamma_1$ 		& $2.80_{-0.27}^{+0.14}$ \\
$\Gamma_2$ 		& $-1.38_{-0.10}^{+0.27}$ \\
$\alpha$ 		& $(3.1_{-1.8}^{+3.2}) \times 10^{-4}$  \\
$E_{fold}$ (keV) 	& $5.54_{-0.13}^{+0.16}$  \\ 
Flux\footnote{\eflux, 2--10\,keV}	& $5.5 \times 10^{-9}$ \\ \hline
			  
Cyclotron line 		&  \\ \cline{1-1}		    
$\rm \tau_{cyc} $ 		&  $0.33_{-0.06}^{+0.07}$  \\ 
$\rm E_{cyc} $ (keV) 	&  $36.2_{-0.7}^{+0.5}$ \\
$\sigma_{cyc}$ (keV) 	&  $3.37_{-0.75}^{+0.92}$  \\ \hline 

$\chi^2_r$/degrees of freedom & 0.97/49 \\ \hline \hline 

\end{tabular}
\end{minipage}
\end{table*}

A clear correlation exists between the line energy and spectral
cut-off indicating that the cut-off is related to the magnetic field.
\citet{Mak92} and \citet{Mak99} first noted this relationship by
plotting cyclotron line energies from \ginga\ and other instruments,
derived with a variety of continuum models, against the PLCUT cut-off
energy from sometimes non-contemporaneous \ginga\ observations.  They
found that the relationship was consistent with a power law, $E_{cyc}
\propto E_{cut}^{1.4}$ (or, $E_{cut} \propto E_{cyc}^{0.7}$),
indicating a saturation as compared to a linear
correlation. Figure~\ref{f:ecutvsecyc} has the advantage that each
point is derived from a uniform model fit to a single spectrum.  And,
with the exception of A0535+26, all points are from the same set of
instruments.  \srcnm\ fits nicely on the correlation, and below
30\,keV, the slope is consistent with $E_{cut} \propto
E_{cyc}^{0.7}$. However, there appears to be a break in the
relationship above \aprx30\,keV.  This flattening
is more abrupt than the smooth turnover of the
$E_{cyc}^{0.7}$ power law suggesting that the processes that form
the continuum saturate at higher magnetic fields.

With \srcnm, we have added a 13$^{th}$ accreting pulsar to the list of
objects with secure cyclotron line detections.  Nearly all of these
have been confirmed, and several were discovered, with \rxte\ and \sax.
We have now begun detailed studies
of these objects as a class. In particular, since ten of the thirteen
sources have been observed with \rxte, we are applying uniform
analyses to all these objects to further understand the line forming
regions.  First results of such studies, including correlations
between other line parameters, are given in \citet{Cob01b}.

\acknowledgements

ASM data are provided by the \rxte/ASM teams at
MIT and at the \rxte\ SOF and GOF at NASA's GSFC.  This work was
supported by NASA grant NAS5-30720, NSF travel grant INT-9815741, 
DFG grant Sta~173/25-1, and a travel grant from the DAAD.


\newpage
\begin{thebibliography}{}

\bibitem[\protect\astroncite{{Araya-G{\'o}chez} \& {Harding}}{2000}]{Ara00}
{Araya-G{\'o}chez}, R.~A., \& {Harding}, A.~K.,  2000, ApJ, 544, 1067


\bibitem[\protect\astroncite{Campana, Israel \& Stella}{1999}]{Cam99}
Campana, S., Israel, G., \& Stella, L.,  1999, A\&A, 352, L91

\bibitem[\protect\astroncite{{Campana} et~al.}{1998}]{Cam98}
{Campana}, S., {Israel}, G.~L., {Stella}, L., \& {Santangelo}, A.,  1998, IAU
  Circ., 7039, 2

\bibitem[\protect\astroncite{Coburn}{2001}]{Cob01b}
Coburn, W.,  2001,
\newblock {\em Ph.D. thesis\/}, University of California, San Diego,
\newblock in preparation

\bibitem[\protect\astroncite{Coburn et~al.}{2001}]{Cob01}
Coburn, W., Heindl, W., Gruber, D., Rothschild, R., Staubert, R., Wilms, J., \&
  Kreykenbohm, I.,  2001, ApJ, 552, 738

\bibitem[\protect\astroncite{{Dal Fiume} et~al.}{2000}]{Dal00}
{Dal Fiume}, D., et~al., 2000,
\newblock in The $\rm 5^{th}$ Compton Symposium, ed. M. McConnell, J. Ryan,
  (Melville, New York: AIP),  183

\bibitem[\protect\astroncite{Heindl et~al.}{2000}]{Hei00}
Heindl, W., et~al., 2000,
\newblock in The $\rm 5^{th}$ Compton Symposium, ed. M. McConnell, J. Ryan,
  (Melville, New York: AIP),  173

\bibitem[\protect\astroncite{{Heindl} et~al.}{1999}]{Hei99}
{Heindl}, W.~A., {Coburn}, W., {Gruber}, D.~E., {Pelling}, M.~R., {Rothschild},
  R.~E., {Wilms}, J., {Pottschmidt}, K., \& {Staubert}, R.,  1999, ApJ, 521,
  L49

\bibitem[\protect\astroncite{Jahoda}{2000}]{Jah00}
Jahoda, K.,  2000, in \emph{Rossi2000}: Astrophysics with the Rossi X-ray
  Timing Explorer

\bibitem[\protect\astroncite{Jahoda et~al.}{1996}]{Jah96}
Jahoda, K., Swank, J.~H., Giles, A.~B., Stark, M.~J., Strohmayer, T., \& Zhang,
  W.,  1996, SPIE, 2808, 59

\bibitem[\protect\astroncite{{Kendziorra} et~al.}{1994}]{Ken94}
{Kendziorra}, E., et~al., 1994, \aap, 291, L31

\bibitem[\protect\astroncite{Knight}{1982}]{Kni82}
Knight, F.,  1982, ApJ, 260, 538

\bibitem[\protect\astroncite{Kretschmar et~al.}{2000}]{Kre00}
Kretschmar, P., Araya-G{\'o}chez, R.~A., Kreykenbohm, I., Wilms, J., Staubert, R.,
  Heindl, W.~A., Rothschild, R.~E., \& Gruber, D.~E.,  2000,
\newblock in Proc. 4th INTEGRAL Symposium,  ESA-SP),
\newblock submitted

\bibitem[\protect\astroncite{{Kreykenbohm} et~al.}{1999}]{Kre99}
{Kreykenbohm}, I., {Kretschmar}, P., {Wilms}, J., {Staubert}, R., {Kendziorra},
  E., {Gruber}, D.~E., {Heindl}, W.~A., \& {Rothschild}, R.~E.,  1999, A\&A,
  341, 141

\bibitem[\protect\astroncite{Liu, {van Paradijs} \& {van den Heuvel}}{2000}]{Liu00}
{Liu}, Q.~Z., {van Paradijs}, J., \& {van den Heuvel}, E.~P.~J., 2000,
A\&AS, 147, 25

\bibitem[\protect\astroncite{Makishima \& Mihara}{1992}]{Mak92}
Makishima, K., \& Mihara, T.,  1992,
\newblock in Frontiers of X-Ray Astronomy (Proc. of the 28$\rm^{th}$ Yamada
  Conf.),  (Tokyo: Uni. Acad. Press), ~23

\bibitem[\protect\astroncite{{Makishima} et~al.}{1999}]{Mak99}
{Makishima}, K., {Mihara}, T., {Nagase}, F., \& {Tanaka}, Y.,  1999, ApJ, 525,
  978

\bibitem[\protect\astroncite{Mihara}{1995}]{Mih95}
Mihara, T.,  1995,
\newblock {\em Ph.D. thesis\/}, University of Tokyo

\bibitem[\protect\astroncite{Rothschild et~al.}{1998}]{Rot98}
Rothschild, R., et~al., 1998, ApJ, 496, 538

\bibitem[\protect\astroncite{Santangelo et~al.}{2001}]{San01}
Santangelo, A., et~al., 2001, ApJ, submitted

\bibitem[\protect\astroncite{{Santangelo} et~al.}{1999}]{San99}
{Santangelo}, A., et~al., 1999, ApJ, 523, L85

\bibitem[\protect\astroncite{Smith \& Takeshima}{1998}]{Smi98}
Smith, D., \& Takeshima, T.,  1998, IAU Circ.,  No. 7014

\bibitem[\protect\astroncite{Tanaka}{1986}]{Tan86}
Tanaka, Y.,  1986,
\newblock in Radiations hydrodynamics in stars and compact objects, ed. D.
  Mihalas, K. Winkler,  (New York, Heidelberg: Springer),  198

\bibitem[\protect\astroncite{White, Swank \& Holt}{1983}]{Whi83}
White, N., Swank, J., \& Holt, S.,  1983, ApJ, 270, 711

\bibitem[\protect\astroncite{Wilms et~al.}{1999}]{Wil99}
Wilms, J., Nowak, M., Dove, J., Fender, R., \& {di Matteo}, T.,  1999, ApJ, 522

\bibitem[\protect\astroncite{Wilson et~al.}{1998}]{Wil98}
Wilson, C.~A., Finger, M.~H., Wilson, R.~B., \& Scott, D.~M.,  1998, IAU Circ.,
   No. 7014

\end{thebibliography}

\end{document}